\def\di{\displaystyle}
\def\&{&\di}
\def\bg{\begin{eqnarray}\begin{array}{rcl}\displaystyle}
\def\bm#1{\begin{eqnarray}\begin{array}{#1}}
\def\eg{\end{array} &\di    &\di   \end{eqnarray}}
\def\bgo{\begin{eqnarray*}\begin{array}{rcl}\displaystyle}
\def\ego{\end{array} &\di    &\di \nonumber  \end{eqnarray*}}
\def\btensor#1#2{\renew\left#1\begin{array}{#2}\di}
\def\etensor#1{\end{array}\right#1}
\documentstyle[12pt]{article}

\def\p{\phi}

\def\m{\mu}

\def\h{\hbox{${1\over2}$}}
\def\q{\hbox{${1\over4}$}}
\def\ei{\hbox{${1\over8}$}}

\def\bib{\bibitem}

\def\e{\epsilon}

\def\p{\phi}
\def\P{\Phi}
\def\h{\hbox{${1\over2}$}}
\def\ex{e^{\h(\P(x)-\p(x))}}
\def\ey{e^{\h(\P(y)-\p(y))}}
\def\e{\epsilon}
\def\B{{\bar K}}
\def\stuff{(\P'-\p')^2+2e^{\h(\P+\p)}+-2e^\P}
\def\F{{\cal F}}
\def\b{\beta}

\def\q{\hbox{${1\over4}$}}
\def\m{\mu}
\def\n{\nu}
\def\pa{\partial}

\def\ei{\hbox{${1\over8}$}}

\def\ps{\p_{ps}}
\def\pp{\pi_{ps}}

\def\beq{\begin{equation}}
\def\eeq{\end{equation}}
\def\bed{\begin{displaymath}}
\def\eed{\end{displaymath}}
\def\beqq{\begin{eqnarray}}
\def\eeqq{\end{eqnarray}}
\def\bedd{\begin{eqnarray*}}
\def\eedd{\end{eqnarray*}}

\def\rene{\renewcommand{\arraystretch}{1.6}}
\def\renew{\renewcommand{\arraystretch}{1}}
\rene
\begin{document}
\begin{flushright}
{\bf
FSUJ-TPI-09/97 \\
DIAS-STP-14/97\\
October 1997\\
}    
\end{flushright}

\par
\vskip .5 truecm
\large \centerline{\Large{\bf On the Canonical Equivalence of}}
\large \centerline{\Large{\bf
    Liouville and Free Fields }}
 
\par
\vskip 1 truecm
\normalsize
\begin{center}
{\bf C.~Ford}\footnote{e-mail: cfo@tpi.uni-jena.de}\\
\sl{Theor.--Phys. Institut, Universit\"at Jena\\ 
Fr\"obelstieg 1\\
D--07743 Jena\\
Germany}
\vskip .5 truecm
{\bf I.~Sachs}\footnote{Present address: Department of Mathematical Sciences, 
University of Durham, UK
}\\
\sl{Dublin Institute for Advanced Studies\\
10 Burlington Road\\
Dublin 4\\
Ireland}
\end{center}
 \par
\vskip 2 truecm
\normalsize

\begin{abstract} 
We obtain the parity invariant generating functional for the canonical
 transformation mapping 
the Liouville theory into a free scalar field and explain how it is related 
to the pseudoscalar transformation
\end{abstract}
\clearpage

In this paper we consider the $2$-dimensional Liouville theory, 
based on the Lagrangian
\bg
{\cal L}=\frac{1}{2}(\partial_\m\P)^2-\frac{m^2}{\b^2}e^{\b\P}.
\eg 
This theory, which describes 
the (world-sheet) gravitational sector of of non-critical strings,  has 
been constructed at the quantum level 
\cite{bct,dj,ow,GN} in the case that 
the ``coupling'' $\b$ is less than $(4\pi)/\hbar$. The construction is based 
on a \sl classical \rm canonical transformation (CT) which maps
Liouville theory to a massless free field. One then proceeds to ``quantise''
the CT, in the sense that equal-time commutators rather than Poisson
brackets are preserved. However, even at the classical level this CT
is not unique.

 Starting with the usual B\"acklund transformation 
Braaten, Curtright and Thorne \cite{bct} gave a time-independent CT 
which maps the Liouville field into a free pseudo scalar field.
Although this transformation is non-local in position space, 
it was demonstrated that the CT  may be derived from the 
remarkably simple generating functional
 \bg\label{standard}
\F=\int^L_{0} dx\left[
\p'\Phi-\frac{\sqrt{8m^2}}{\b^2}e^{\h\b \Phi}\sinh(\h \b
\p)\right].\eg  
As expected, the generating functional (\ref{standard}) is not 
parity-invariant. This manifests itself as a pseudo scalar conformal 
improvement term in the free theory Hamiltonian density. 
A parity invariant CT exchanging the Liouville
and free fields has been given by D'Hoker and Jackiw \cite{dj}. 
This transformation, originally formulated on the light cone was 
later obtained in terms of the usual
canonical phase space variables by Otto and Weigt \cite{ow}. 
Although this CT has been studied  by many
authors (eg. see ref. \cite{FIT} for a recent 
investigation) to our knowledge
the generating functional for the Liouville $\rightarrow$
scalar CT has not been given before.

In this letter we obtain the parity invariant generating functional 
for the Liouville $\rightarrow$ scalar  CT. This new generating functional 
is more complicated than that for the pseudo-scalar field
(\ref{standard}),
yet like (\ref{standard}) it is a \sl local \rm functional of
the Liouville and free fields. 
We furthermore determine the CT between the free and pseudo-scalar fields and 
find that in addition to exchanging space- and time derivatives it, too 
involves screening charges, showing that the scalar- and pseudo scalar 
approaches are equivalent only up to screening charges. 
This is the classical analogue of the same result obtained earlier by 
Gervais and Schnittger \cite{GS} for the quantum Liouville exponentials. 
For completeness, we also give the generating functional 
for the scalar $\rightarrow$  pseudo-scalar transformation.

First we briefly recall the D'Hoker Jackiw CT, and its reconstruction 
in terms of phase space variables. 
Let $\p=\p^+(x^+)+\p^-(x^-)$ be a free field satisfying the
Poisson bracket relations
\bg
\{\p^-(w^-),\p^-(z^-)\}&=&\q \e(w^--z^-), \\
\{\p^+(w^+),\p^+(z^+)\}&=&\q \e(w^+-z^+), \\
\{\p^+(w^+),\p^-(z^-)\}&=&0, 
\eg
$w\pm$ being light cone coordinates, and $\epsilon(w)$ is the sign
function.
Then using these brackets one can show that transformation \cite{dj}
\bg
\label{jackiw}
\P=\p-\frac{2}{\beta}\log\left[1+\h m^2 \Box^{-1}e^{\b\p}\right],\eg
satisfies
\bg
\left.\{\P(w),\P(z)\}\right|_{w^+=z^+}&=&\q \e(w^--z^-),\\
\left.\{\P(w),\P(z)\}\right|_{w^-=z^-}&=&\q \e(w^+-z^+),
\eg
and makes the (improved) Liouville $\P$-field's energy-momentum tensor
 take on the free (improved) form, ie. 
$\Theta_{\mu \nu}^{Liouville}=\Theta_{\mu\nu}^{
Free}$, where
\bg
\Theta_{\mu\nu}^{Liouville}&=\&
\pa_\m \P\pa_\n\P-\frac{1}{2}g_{\m\n}\pa_\alpha\P\pa^\alpha\P-
\beta^{-2}g_{\m\n}m^2e^{\b\P}+2\beta^{-1}(g_{\m\n}\Box-\pa_\m\pa_\n)\P, \\
\Theta_{\mu\nu}^{Free}&=\&
\pa_\m\p\pa_\n\p-\frac{1}{2}g_{\m\n}\pa_\alpha\p\pa^\alpha\p
               +2\beta^{-1}(g_{\m\n}\Box-\pa_\m\pa_\n)\p.
\eg

We now express this CT in an explicitly time-independent form using 
 phase space variables
\bg\label{trafo}
\P(x)&=\& \p(x)-\frac{2}{\b}\log\left[1-\ei K(x)\bar K(x)\right],\\
\Pi(x)&=\&\pi(x)+\frac{1}{4\b}
\frac{K'(x)\bar K(x)-K(x){\bar K}'(x)}{
1-\ei K(x)\bar K(x)}.
\eg
Here
\bg\label{bigk}
K(x)&=\&\frac{1}{2\sinh(\q\b P)}\int^L_{0}dx'
      \exp\left[\frac{\b P}{4}\e(x-x')\right]
k(x'),\\
\bar K(x)&=\&-\frac{1}{2\sinh(\q\b P)}\int^L_{0}dx'
 \exp\left[-\frac{\b P}{4}\e(x-x')\right]
\bar k(x'),\eg
with
\bg
k(x)&=\&m
\exp\left[\frac{\b}{2}\phi(x)+
\frac{\b}{4}\int^L_{0}dz\,\e(x-z)
\pi(z)\right],\\
\bar k(x)&=\&m
\exp\left[\frac{\b}{2}\phi(x)
-\frac{\b}{4}\int^L_{0}dz\,\e(x-z)
\pi(z)\right],
\eg
where
\bg P=\int^L_{0}dz\,\pi(z),\eg
is the zero mode of $\pi(x)$, and $\epsilon(x)$ is the stair step function
\bg\epsilon(x)=2n+1\quad
\hbox{for}\quad
nL<x<(n+1)L,\eg
which coincides with the sign function for $-L<x<L$.
Note that
\bg
K'(x)=k(x)\quad\hbox{and}\quad
\bar K'(x)=\bar k(x).\eg 
$K(x)$ and $\bar K(x)$ are screening charges (ie. of conformal dimension zero). 
Their role played in the construction of the Liouville exponential 
has been emphasised in \cite{FIT,GS}. With the notation chosen here 
it may not be immediately obvious that this CT coincides with that given 
by Otto and Weigt \cite{ow} but one can show that they are indeed identical. 

Although the free field $\phi(x)$ and its conjugate momentum $\pi(x)$ are
assumed to be periodic, ie.
\bg
\phi(x+L)=\phi(x),\quad
\pi(x+L)=\pi(x),\eg
the exponentials $k(x)$ and $\bar k(x)$ are not
\bg
k(x+L)=\exp\left[\frac{\b P}{2}\right]k(x),\quad
\bar k(x+L)=\exp\left[-\frac{\b P}{2}\right]\bar k(x).
\eg
>From (\ref{bigk}) one can see that $K(x)$ and $\bar K(x)$ have
the same periodicity properties as $k(x)$ and $\bar k(x)$,
respectively. In particular the combinations $K(x)\bar K(x)$, 
$K(x)\bar K'(x)$,
$K'(x)\bar K(x)$ and hence $\Phi(x)$ and $\Pi(x)$ are
periodic. Under (\ref{trafo}) the Liouville Hamiltonian density 
\bg\label{ham}
{\cal H}^{Liouville}=
\frac{1}{2}\Pi^2+\frac{1}{2}{\P'}^2+\frac{m^2}{\beta^2}e^{\b\P}-{2\over\b}\P'',
\eg
is then mapped into the (improved) free field form
\bg\label{ham2}
{\cal H}^{Free}={1\over2}\pi^2+{1\over2}{\p'}^2-{2\over\b}\p''.
\eg
Since the transformation is canonical we have 
\bg
\int^{L}_{0}dx\left(
\pi(x) d\p(x)-{\cal H}^{Free}(x)dt\right)
=
\int^{L}_{0}dx\left(
\Pi(x) d\P(x)-{\cal H}^{Liouville}(x)dt\right)
+d{\cal F},\nonumber\eg
where $\F$ is the generating functional of the CT, that is
\bg\label{1aa}\pi(x)={\delta\F\over{\delta\p(x)}},\quad
\Pi(x)=-{\delta\F\over{\delta\Phi(x)}}.\eg
To construct ${\cal F}$, we first express $\pi$ and $\Pi$ in terms of the 
fields $\phi$ and $\Phi$. Differentiating the first of 
eqs. (\ref{trafo}) twice yields
\def\Pp{{\Psi_+}}
\def\Pm{{\Psi_-}}
\def\h{\hbox{${1\over2}$}}
\def\ex{e^{\h(\P(x)-\p(x))}}
\def\ey{e^{\h(\P(y)-\p(y))}}
\def\e{\epsilon}
\def\B{{\bar K}}
\def\guff{4{\Pm'}^2-2m^2(e^\Pm-1)e^\Pp}
\def\stuff{(\P'-\p')^2+2\b^{-2}m^2\left(e^{\h\b(\P+\p)}-e^{\b\P}\right)}
\def\stufftwo{\nabla(\P-\p)\cdot\nabla(\P-\p)+
2\beta^{-2}m^2\left(e^{\h\b(\P+\p)}-e^{\b\P}\right)}
\def\F{{\cal F}}
\bg\label{te}
\P''&=\&\p'' +\h(\P'-\p')^2+
{\p'(K'\B+K\B')+\pi(K'\B-K\B')+4\beta^{-1}K'\B'\over{8(1-\ei K\B)}}, \\
&=\&\p''+\h\beta (\P'-\p')^2+\h\beta \p'(\P'-\p')+\h\beta\pi(\Pi-\pi)+
\h \b^{-1}m^2 e^{\h\b (\P+\p)}.
\eg
Using (\ref{te}) and ${\cal H}^{Liouville}={\cal H}^{Free}$, we
can write the momenta in terms of $\phi$ and $\Phi$ as follows  
\bg\label{type1a}\h \pi={\b^{-1}(\P''-\p'')-\h{\P'}^2+\h\P'\p'-\h
  \b^{-2}
m^2e^{\h\b(\P+\p)}\over{
\sqrt{\stuff}}},\eg
\bg\label{type1b}
\h\Pi={\b^{-1}(\P''-\p'')+\h{\p'}^2-\h\p'\P'-\beta^{-2}m^2e^{\b\P}+\h 
\b^{-2}m^2e^{\h\b(\P+\p)}\over
{\sqrt{\stuff}}}.
\eg
It is convenient to introduce the following variables
\bg\Pm=\h\beta(\P-\p),\qquad \Pp=\h\beta(\P+\p).\eg
Eqs. (\ref{1aa}) then become 
\bg
{\delta\F\over{\delta\Pp(x)}}\&=\&\frac{1}{\beta}\left(
{\delta\F\over{\delta\p(x)}}+{\delta\F\over{\delta\P(x)}}\right)=
\frac{\pi-\Pi}{\beta}=-\b^{-2}\sqrt{
4{\Pm'}^2-2m^2(e^\Pm-1)e^\Pp},\\\nonumber \di
{\delta\F\over{\delta\Pm(x)}}\&=\&
\frac{1}{\beta}\left({\delta\F\over{\delta\P(x)}}-{\delta\F\over{\delta\p(x)}}\right)
=
-\frac{\Pi+\pi}{\beta}=-
{8\Pm''-4\Pm'\Pp'-2m^2e^{\Pm+\Pp}\over{\b^{2}\sqrt{\guff}}}\nonumber.
\eg
\beq
\label{equations}
\eeq
Now the crucial point is that the $\delta \F/\delta\Pp$
equation can be integrated at once to give 
\bg\label{integral} \F&=\&-\frac{1}{\b^2}\int^L_0 dx
\int^\Pp d\Psi\sqrt{
4{\Pm'}^2-2m^2(e^\Pm-1)e^{\Psi}}+K[\Pm]\\
&=\&\frac{1}{\b^2}\int^L_0 dx \Bigl[-2\sqrt{\guff}\\&\&+
\quad \left.
2\Pm'\log\left(
{2\Pm'+\sqrt{\guff}\over{2\Pm'-\sqrt{\guff}}}\right)\right]
+K[\Pm],\eg
where $K[\Pm]$ is some functional of $\Pm$.
It is straightforward to show  that equation (\ref{integral})
is consistent with the $\delta\F/\delta\Pm$ equation
in (\ref{equations}) if $K[\Pm]=0$. Finally, substituting again the 
original variables we  obtain the  
generating functional for the D'Hoker Jackiw CT
\bg\label{result}
\F&=\&\frac{1}{\b}\int^L_0  dx\Bigl[-2\sqrt{\stuff}\\
&+\& \left. (\P'-\p')\log\left(
\frac{\P'-\p'+\sqrt{\stuff}}{\P'-\p'-\sqrt{\stuff}}
\right)
\right],\eg
which is our main result.

Recall that (\ref{result}) generates the CT 
mapping the Liouville field to the free scalar field whereas  (\ref{standard}) 
maps ${\cal H}^{Liouville}$ to ${\cal
H}^{Free}$
with  a \sl pseudo-scalar \rm  improvement term, ie.
\bg{\cal H}^{Free}={1\over2} \pi^2+{1\over2} \phi'^2+{2\over\beta}
\pi'.\eg
Thus we have two  CT's relating the Liouville theory to distinct
free theories. Hence, there is a CT which
exchanges the two free theories. Let us now 
construct this transformation in turn. Using the generating 
functional (\ref{standard}) one has the following
relation between
the pseudo-scalar variables $(\ps,\pp)$ and the Liouville variables
$(\P,\Pi)$
\bg
\pp&=\&{\delta {\cal F}\over{\delta\ps}}=-\P'-\frac{\sqrt{2 m^2}}{\beta}e^{\h\b\P}
\cosh\left(\h\b\ps\right),\\
\Pi&=\&-{\delta {\cal F}\over{\delta \P}}=-\ps'+
\frac{\sqrt{2m^2}}{\b}e^{\h\b\P}\sinh\left(\h\b\ps\right).\eg
Writing the Liouville variables in terms of the free
fields $(\p,\pi)$ yields
\bg\label{ps}
\pp&=\&-\p'-\frac{1}{4\b}{K'\bar K+K\bar K'\over{1-\ei K\bar K}}-
{\sqrt{2m^2}e^{\h\b\p}\over{\b(1-\ei K\bar
    K)}}\cosh\left(\h\b\ps\right),\\ \di
\ps'\&=\&-\pi-\frac{1}{4\b}{K'\bar K-K\bar K '\over{1-\ei K\bar K}}
+\frac{\sqrt{2m^2}e^{\h\b\p}}{{\b(1-\ei K\bar
    K)}}\sinh\left(\h\b\ps\right).\eg
Now, if one takes the \sl negative \rm square root of $2m^2$, it
is easy to see that (\ref{ps}) has the solution
\bg\label{psct}
\ps(x)&=\&-\h\int^L_0 dx'\e(x-x')\pi(x') +\frac{2}{\b}\log\left(K(x)/\sqrt{8}\right),\\
 \pp(x)&=\&-\p'+\frac{2}{\b}\frac{K'(x)}{K(x)}.\eg
Similarly, if one takes the \sl positive \rm square root of $2m^2$,
we have 
\bg\label{cff}
\ps(x)&=\&-\h\int^L_0 \e(x-x')\pi(x')-\frac{2}{\b}\log\left(\bar K(x)/\sqrt{8}\right),\\
\pp(x)&=\&-\p'(x)+\frac{2}{\b}\frac{\bar K'(x)}{\bar K(x)}.\eg
Note the presence of the screening charges in (\ref{psct}). This may come as 
a surprise, because after all we are mapping a free field into a free field 
and for that the transformation obtained from (\ref{psct}) by dropping 
these terms would be the more obvious candidate. 
This was known before to be true at the 
operator level. The present result shows that this can be seen even at the 
classical level. 

Finally, we  derive the generating functional 
for (\ref{psct}).
To find this we use
\bg K'(x)=me^{\h\b\p-\h\b\ps+\log(K/\sqrt{8})}=
{mK\over{\sqrt{8}}}e^{\h\b(\p-\ps)},\eg
and so
\bg{K'(x)\over{K(x)}}={m\over{\sqrt{8}}}e^{\h\b(\p-\ps)}.\eg
Thus the CT (\ref{psct}) can be written as
\bg
\pp&=\&-\p'+{m\over{\sqrt{2}}\b}e^{\h\b(\p-\ps)},\\
\pi&=\&-\ps'+{m\over{\sqrt{2}}\b}e^{\h\b(\p-\ps)}.\eg
This transformation is generated by
\bg{\cal F}(\ps,\p)=
\int_0^L dx\left[
-\ps \p'-\frac{\sqrt{2} m}{\b^2}e^{\h\b(\p-\ps)}\right].\eg
Similarly, one can obtain the generating functional for
the CT (\ref{cff}).

\section*{Acknowledgements}
 We thank G. Weigt for a 
helpful suggestion and A. Wipf for general discussions on canonical
transformations.
C. F thanks the DFG for financial support.

\section*{Note Added (May 1999)}

\tt
It has recently come to our attention that the generating functional
(24) was obtained by Jordadze \cite{george} in the early 1980's.
We therefore feel it  important to set the record straight. 
C. F. thanks George Jordadze for the friendly and constructive way in which he
brought this matter to our attention.

\rm

\end{document}